\documentstyle[12pt,amssymb,a4,psfig]{article}
\newcommand{\eem}[1]{#1_{\mbox {\small eem}}}
\newcommand{\vmd}[1]{#1_{\mbox {\small vmd}}}
\newcommand{\vI}{V_{\mbox {\scriptsize I}}}
\newcommand{\vII}{V_{\mbox {\scriptsize II}}}

\begin{document}

\begin{center}
{\Large {\bf
A CONSISTENT EXPLANATION OF THE ROPER PHENOMENOLOGY
}}
\end{center}

\vspace{1cm}

\begin{center}
F. Cano and  P. Gonz\'alez
\end{center}

\begin{center}
Departamento de F\'{\i}sica Te\'orica and IFIC,\\
Centro Mixto Universidad de Valencia-CSIC\\
46100 Burjassot (Valencia), Spain
\end{center}

\vspace{3cm}

\begin{abstract}
{\small We study the electromagnetic transitions of the Roper
$N(1440)$ resonance. Our results, when combined with the previously
obtained for the mass and the pionic strong decay widths of the Roper,
show that within a non-relativistic constituent quark model
scheme, a comprehensible understanding of the Roper 
phenomenology can be achieved. They also seem to support the view
of the Roper as a radial excitation of the nucleon, though more 
experimental data are needed to reach a definitive conclusion.}
\end{abstract}

\vfill

\vspace{3cm}
\noindent 
cano@xaloc.ific.uv.es \\
gonzalep@evalvx.ific.uv.es

\vfill
 
\newpage

	Baryon resonances and in particular the Roper resonance
N(1440) deserve special attention at 
the current moment from the theoretical as well as the experimental
point of view. The Roper resonance has been observed at Saturne as an
excitation of a hydrogen target scattered by $\alpha$-particles
\cite{MORSCH92}. The transition form factors for the excitation of
baryon resonances are going to be systematically measured at TJNAF in
a near future. Related experimental programs are carried out in MAMI,
ELSA and GRAAL. Theoretically, to predict the precise position of the
Roper in the nucleon resonance spectrum and the correct values (including
signs) of its strong and electromagnetic transition amplitudes has
been a challenge for many years. 

	Though the simplest image of the Roper corresponds to a radial
excitation of the $3q$ nucleon state, due to the
difficulty in describing its phenomenology, 'more completed' or
alternative descriptions have been proposed. Among the first ones we
can mention the relativized versions of the  constituent
quark model \cite{CAPSTICK92,CLOSE90} and the light front constituent
quark model \cite{CARDARELLI97,CAPSTICK95}.
Among the second ones the breathing bag model
\cite{BROWN83, GUICHON85} and the hybrid state descriptions
\cite{BARNES83,CLOSE88,CARLSON91,LI92} which confer to the Roper an exotic
nature. Nonetheless a unified description of the Roper phenomenology
has not been achieved and its nature is still under discussion. 

	In the last few years very precise fits of the low-lying
baryon spectrum have been obtained 
making use of potential models \cite{DESPLANQUES92,DONG94,GLOZMAN96}. 
One of the models \cite{DESPLANQUES92}
has also been succesfully applied to predict the pionic decay widths
of the low-lying nucleon and delta resonances \cite{CANO96}.
Concerning the Roper, the consideration, via a $^3P_0$ model, of the
$q\bar{q}$ pion structure, becomes essential to correctly reproduce the
strong pionic decay data. 
The purpose of this paper is to apply the same model to the
calculation of photo and electroproduction amplitudes of the Roper,
the aim being to check whether it is possible or not, from a
non-relativistic constituent quark model, to get a unified description
of the Roper phenomenology and to learn about its elusive nature. 

\section{Photo and electroproduction amplitudes}

	Some years ago, Gavela et al. \cite{GAVELA80} got a reasonable
description at that time of the Roper photoproduction amplitude
$N(1440) \rightarrow N \gamma$ by combining vector meson dominance (VMD)
with a $^3P_0$ model for the meson production. A simple and immediate 
generalization to $Q^2 \neq 0$ is obtained by introducing an
intermediate propagator for the meson (a more fundamental, but
at the same time more complicated technical approach would be to
couple the photon to the $q\bar{q}$ structure of the meson).
The transition matrix element for a general process $B \rightarrow B'
\gamma$  reads:

\begin{equation}
\label{vmdsakuraiq}
\langle B' \gamma | H | B \rangle = \sum_V \langle B' V | H_{^3P_0} | B
\rangle \frac{1}{m_V^2 + Q^2} \langle \gamma | H_{V\gamma} | V \rangle
\end{equation}

\noindent
with $V=\rho$, $\omega$ and \cite{SAKURAI69}:

\begin{equation}
\label{couplingsakurai}
\langle \gamma | H_{V\gamma} | V \rangle =  e \frac{m_{V}^2}{f_V} 
\end{equation}

	The matrix element in the $^3P_0$ model is given by
\cite{LEYAOUANC73}:

\begin{eqnarray}
\label{imbbv}
\langle B' V | H_{^3P_0}| B \rangle & = & - 3 \gamma \sum_m (110 | m\;
-m) \nonumber \\
& &  \int d\vec{p}_1 \;  d\vec{p}_2 \; d\vec{p}_3
\; d\vec{p}_4 \; d\vec{p}_5 \; {\cal{Y}}_{1m} (\vec{p}_4 - \vec{p}_5)
\delta(\vec{p}_4 + \vec{p}_5) \Phi_{\mbox {\tiny Pair}}^{-m} \nonumber
\\ & & \left[ \Psi_{B'}(\vec{p}_1,\vec{p}_2,\vec{p}_4)
\Phi_{B'}\right]^{*} \left[ \Psi_{V}(\vec{p}_3,\vec{p}_5)
\Phi_{V}\right]^{*} \left[ \Psi_{B}(\vec{p}_1,\vec{p}_2,\vec{p}_3)
\Phi_{B}\right] \nonumber \\
 & & 
\end{eqnarray}

\noindent
where $\gamma$ is the pair creation coupling constant, ${\cal{Y}}_{1m}$ a
solid harmonic, $\Psi$ stands for the momentum space wave function and
$\Phi$ for the spin-isospin wave function.

	By using eqs. (\ref{couplingsakurai}) and (\ref{imbbv}) we can
write after some algebra the transition matrix element as:

\begin{eqnarray}
\langle B' \; \gamma | H | B \rangle & = & - 3 e \gamma 
\left(\frac{m_A}{\omega_\gamma}\right)^{1/2}
\delta^{(3)}(\vec{P}_B - \vec{P}_B \,' - \vec{q}\,) \left( -
\frac{1}{\sqrt{4 \pi}} \frac{1}{2} \sqrt{\frac{2}{3}} \right)
\nonumber \\
& & \left[ \frac{1}{f_\omega} \frac{m_\omega^2}{m_\omega^2 + Q^2} 
\langle \eta_{B'} | \eta_B \rangle  +  \frac{1}{f_\rho}
\frac{m_\rho^2}{m_\rho^2 + Q^2} \langle \eta_{B'} |
\tau_0^{(3)} | \eta_B \rangle \right] \nonumber \\
 & & \langle \chi_{B'} | \int 
d\vec{p}_{\xi_{1}} \; d\vec{p}_{\xi_{2}} \; 
\Psi_{B'}^{*}(\vec{p}_{\xi_{1}},\vec{p}_{\xi_{2}} +
\sqrt{\frac{2}{3}} \vec{q}\,) 
\Psi_{V}^{*}(-\sqrt{\frac{2}{3}} \vec{p}_{\xi_{2}} + \frac{\vec{P}_B}{3} -
\frac{\vec{q}}{2}) \nonumber \\
 & & \left[  (\vec{p}_3 - \vec{q}\,) \vec{\epsilon}^{\: *} + i
\vec{\epsilon}^{\: *}
(\vec{\sigma}^{(3)} \times (\vec{p}_3 - \vec{q}\,))
\right] \Psi_{B}(\vec{p}_{\xi_{1}},\vec{p}_{\xi_{2}})
| \chi_{B} \rangle  
\label{metotalvmd}
\end{eqnarray}

\noindent 
where $m_A$ is an average $\rho$ and $\omega$ mass, $\omega_\gamma$
($\vec{q}\,$) is the photon energy (trimomentum), $\eta$ stands for the
isospin wave function, $\chi$ for the spin one,  ($\vec{\xi}_{1}$,
$\vec{\xi}_{2}$) are the Jacobi coordinates and ($\vec{p}_{\xi_{1}}$, 
$\vec{p}_{\xi_{2}}$) its associated momenta operators. The
polarization of the vector meson has been denoted by $\vec{\epsilon}$.
$\vec{\sigma}$'s ($\vec{\tau}\,$'s) as usual denote spin (isospin)
matrices. The superindex in the spin and isospin operators 
indicates the quark on which they act. 

	All the dependence on a specific quark model for the baryon is
contained in the baryon wave functions $\Psi_B$, $\Psi_{B'}$. 
For $\Psi_V$ we shall take a gaussian form \cite{LEYAOUANC73}:

\begin{equation}
\Psi_V(\vec{q}\,) = \frac{2 R_A^{3/2}}{\pi^{1/4}} \exp \left( -
\frac{R_A^2}{2} \vec{q}^{\: 2}\right) Y_{00}(\hat{q})
\end{equation}

 	The value chosen for $R_A$ is 2.82 GeV$^{-1}$ \cite{GAVELA80},
quite close to the one obtained from the study of the leptonic decay
of $\rho^0$ in the quark model. If we assume the SU(3) constraint,
$f_\omega = 3 f_\rho$, the only free parameter is the ratio
$\frac{\gamma}{f_{\rho}}$. 

\subsection{Nucleon form factors.}
	As a first application of expression (\ref{metotalvmd}) it is
interesting to analyze the form factors of the nucleon for which we
use the Breit system of reference. To get the magnetic form factor we
take a transverse polarization vector $\vec{\epsilon}_{\pm} = \mp  
\frac{1}{\sqrt{2}} (1,\pm i, 0)$ and compare expression
(\ref{metotalvmd}) with the corresponding matrix element in terms of
the form factors of the nucleon. Thus we get:

\begin{eqnarray}
G_M^{p,n}(Q^2) & = & - 3 \gamma
(m_A)^{1/2}
\left( -\frac{1}{\sqrt{4 \pi}} \frac{1}{2} \sqrt{\frac{2}{3}} \right)
( 2 M_N (2 \pi)^{3/2} )
\nonumber \\
& & \left[ \frac{1}{f_\omega} \frac{m_\omega^2}{m_\omega^2 +
Q^2}
\langle \eta_{B'}\left(\frac{1}{2}, \pm \frac{1}{2}\right)  |
\eta_B \left(\frac{1}{2}, \pm \frac{1}{2}\right) \rangle \right.
\nonumber \\
 & & \left. +  \frac{1}{f_\rho}
\frac{m_\rho^2}{m_\rho^2 + Q^2} \langle \eta_{B'}
\left(\frac{1}{2}, \pm \frac{1}{2}\right)  |
\tau_0^{(3)} | \eta_B
\left(\frac{1}{2}, \pm \frac{1}{2}\right)  \rangle \right] \nonumber
\\
 & & \langle \chi_{B'} \left(\frac{1}{2}, - \frac{1}{2}\right) | \int
d\vec{p}_{\xi_{1}} \; d\vec{p}_{\xi_{2}} \;
\Psi_{B'}^*(\vec{p}_{\xi_{1}},\vec{p}_{\xi_{2}} +
\sqrt{\frac{2}{3}} \vec{q}\,)
\Psi_{V}^*(-\sqrt{\frac{2}{3}} \vec{p}_{\xi_{2}} -
\frac{\vec{q}}{3}) \nonumber \\
 & & \frac{1}{q} \left[  - q \sigma^{(3)}_{-} + p_{3_z}
\sigma^{(3)}_{-}
- p_{3 -} \sigma^{(3)}_{z} - p_{3 -} \right] 
\Psi_{B}(\vec{p}_{\xi_{1}},\vec{p}_{\xi_{2}})
| \chi_{B} \left(\frac{1}{2}, \frac{1}{2}\right) \rangle
\nonumber \\
 & &
\label{gmvmd}
\end{eqnarray}

\noindent
$M_N$ being the nucleon mass.

	The electric form factor is given in terms of the time
component of the current ($G_E(Q^2) = \langle j_0 \rangle$) that can be
related to the spatial one $j_3$ by gauge invariance ($j_0 =
\frac{q}{\omega} j_3$, $\vec{q} \equiv (0,0,q)$). Then:

\begin{eqnarray}
G_E^{p,n}(Q^2) & = & - 3 \gamma
(m_A)^{1/2}
\left( -\frac{1}{\sqrt{4 \pi}} \frac{1}{2} \sqrt{\frac{2}{3}} \right)
(- \sqrt{2} (2 \pi)^{3/2} )
\nonumber \\
& & \left[ \frac{1}{m_\omega f_\omega} \frac{m_\omega^2}{m_\omega^2 +
Q^2}
\langle \eta_{B'}\left(\frac{1}{2}, \pm \frac{1}{2}\right)  | 
\eta_B \left(\frac{1}{2}, \pm \frac{1}{2}\right) \rangle  \right.
\nonumber \\
 & & +  \left. \frac{1}{m_\rho f_\rho}
\frac{m_\rho^2}{m_\rho^2 + Q^2} \langle \eta_{B'}
\left(\frac{1}{2}, \pm \frac{1}{2}\right)  |
\tau_0^{(3)} | \eta_B
\left(\frac{1}{2}, \pm \frac{1}{2}\right)  \rangle \right] \nonumber
\\
 & & \langle \chi_{B'} | \int
d\vec{p}_{\xi_{1}} \; d\vec{p}_{\xi_{2}} \;
\Psi_{B'}^*(\vec{p}_{\xi_{1}},\vec{p}_{\xi_{2}} +
\sqrt{\frac{2}{3}} \vec{q}\,)
\Psi_{V}^*(-\sqrt{\frac{2}{3}} \vec{p}_{\xi_{2}} -
\frac{\vec{q}}{3}) \nonumber \\
 & & q \left[ p_{3_z} - q + \sigma^{(3)}_{+} p_{3 -} -
\sigma^{(3)}_{-} p_{3 +}
\right] \Psi_{B}(\vec{p}_{\xi_{1}},\vec{p}_{\xi_{2}})
| \chi_{B} \rangle
\label{gevmd}
\end{eqnarray}

	Symbolically 
$\langle \chi_{B'} |\sigma^{(3)}_{-}|\chi_{B}\rangle
  G_E(Q^2) \propto Q^2 G_{M}(Q^2)$. Let us note
that the factor $q$ multiplying the last bracket makes the electric
form factor to vanish at $Q^2 = 0$, what can be interpreted as
an indication that the total charge of the mesonic cloud 
($\rho$, $\omega$) in the nucleon
is zero. Therefore, in order to get the baryon charge for
$G_E^p(Q^2=0)$, VMD can not be the only mechanism for the
photon-baryon interaction.

	We propose instead an extended vector meson dominance model
(EVMD) as schematically pictured in fig. 1. The first diagram on the
r.h.s. corresponds to VMD as calculated above. The second diagram on 
the r.h.s. corresponds to a direct coupling to the baryon for which we
shall use a modified elementary emission model \cite{LEYAOUANC88} 
where a convergent $(p/E)$ instead of a $(p/m)$ expansion has been applied 
\cite{CANO97}. Quite similar assumptions underlie the so called
two-phase models \cite{BROWN86} where the electromagnetic properties
of the nucleon come from two sources, the vector meson cloud and the
core of quarks respectively corresponding to the two diagrams we
consider in fig. 1. Hence we can summarize the EVMD scenario in the
equations:  

\begin{eqnarray}
G_E^{p,n}(Q^2) & = & \eem{[G_E^{p,n}(Q^2)]} + \vmd{[G_E^{p,n}(Q^2)]} \\
G_M^{p,n}(Q^2) & = &  c \eem{[G_M^{p,n}(Q^2)]} + \vmd{[G_M^{p,n}(Q^2)]}
\label{gmevmd}
\end{eqnarray}
 
\noindent
where the relative weight parameter of the two diagrams $c$
plays the same role as an anomalous magnetic moment of the quarks.
Note that although the value of the quark mass used is $m_q \approx
\frac{M_N}{3}$, due to the $(p/E)$ expansion in the EEM, the magnetic
moment is not saturated by $\eem{[G_M^{p,n}(Q^2=0)]}$, the remaining
contribution having to be provided by $\vmd{[G_M^{p,n}(Q^2=0)]}$.
For the electric form factor the relative weight is 1 due to
the charge conservation (recall that $\vmd{[G_E^{p,n}(Q^2=0)]} = 0$). 

	To fix the two parameters $\frac{\gamma}{f_\rho}$ and $c$, we
can use the magnetic moment $\mu_p$ and the square mean radius of the
proton ($\langle r_p^2 \rangle$) for instance. One should be aware 
however that important contributions to $\langle r_p^2 \rangle$, as
the one from the Darwin-Foldy term $ \approx (\frac{2}{3 m_q^2})$,
have not been considered. Actually, a value of $\langle r_p^2 \rangle$
between 0.5 and 0.6 fm$^2$ seems to be a better choice to get a very
good fit of most electromagnetic transitions \cite{CANOPREP}.

	In order to extract some results we shall make use of two
potential models $\vI$, $\vII$, extensively detailed elsewhere
\cite{CANO96}.  
For the sake of completeness we only quote here their expressions 
(values of the parameters as in ref. \cite{CANO96}): 

\begin{eqnarray}
\vI & = & \sum_{i<j} \frac{1}{2} \left[ \frac{r_{ij}}{a^2} - 
\frac{\kappa}{r_{ij}} + \frac{\kappa}{ m_{i} m_{j}} 
\frac{\exp (-r_{ij}/r_{0})}{r_{0}^{2} r_{ij}} \vec{\sigma}_{i} 
\vec{\sigma}_{j}-D
\right] \\ 
\vII  & = &  \sum_{i<j} \frac{1}{2} \left[ \frac{r_{ij}}{a^2} -
\frac{\kappa}{r_{ij}} + \frac{\kappa}{6 m_{i} m_{j}}
\frac{\exp (-r_{ij}/r_{0})}{r_{0}^{2} r_{ij}} \vec{\sigma}_{i}
\vec{\sigma}_{j}-D
\right] \nonumber \\
 & +  &  \sum_{i\neq j \neq k \neq i} \frac{1}{2}
\frac{V_{0}}{m_{i} m_{j} m_{k}} \frac{e^{-m_{0} r_{ij}}}{m_{0} r_{ij}}
\frac{e^{-m_{0} r_{ik}}}{m_{0} r_{ik}}
\end{eqnarray}

$\vI$ contains
the 'minimal' one gluon exchange kind of potential
and provides a reasonable average fit for the meson and baryon spectra
(not fitting the Roper in the spectrum) 
\cite{SILVESTRE85} whereas $\vII$, that incorporates a
phenomenological three--quark interaction, fits very precisely the low lying
baryon spectrum \cite{DESPLANQUES92} and most of its pionic strong 
decay widths, in particular the Roper resonance $N(1440)$ mass and 
decay widths \cite{CANO97}.

	By fixing $\mu_p = 2.79$ and $\langle r_p^2 \rangle = 0.54$
fm$^2$ one gets $c=0.49 (0.17)$ and $\frac{\gamma}{f_\rho} = -0.38
(-0.71)$ for $\vI$ ($\vII$). Then for the neutron we get 
$\mu_n = -1.82$ (-1.84) and
$\langle r_n^2 \rangle = -0.003$ (-0.002) for  $\vI$ ($\vII$). The
separate contributions from the two diagrams are shown in table 1.

	Let us emphasize that we do not do a completely consistent
construction
of the transition operator according to the mass operator. We assume
that the wave functions we obtain by solving the Schr\"odinger equation
for $\vI$ of $\vII$ represent, through the effective parameters and
interactions of the potential a sort of averaged $(p/E)$ quark core wave
function. One should realize however that the corrections introduced
by the $(p/E)$ expansion in the transition operator get correlated to
the model wave functions through the weight parameter.
 
	It is ilustrative to draw the magnetic form factor. 
Looking at fig. 2 it is clear that 
the data lie between the predictions of the two models.
This emphasizes the major role of the form of the 
transition operator (as it was
the case in the strong decays of the Roper where the $q\bar{q}$
structure of the pion was essential) leaving a fine tunning effect to
the specific quark potential model employed. In particular, the
$^3P_0$ structure is mostly responsible to get the tendency of
experimental data.

\subsection{Roper transition amplitudes.}

	The Roper transition amplitudes calculated in the EVMD model  
are shown in figs. 3.a and 3.b. 
Though a clarification of experimental data seems
to be necessary, it is worth to emphasize that we reproduce the
experimental sign of the transverse amplitude at $Q^2=0$ (contrary to
what happens with an EEM calculation) for what the VMD as provided by
the $^3P_0$ model becomes
essential (see table 1). The smallness of the EEM contribution is
related to the use of a $(p/E)$ expansion instead of a $(p/m)$ one,
combined with the presence of a node in the Roper wave function (note
the higher sensitivity of this process as compared to the magnetic
nucleon form factor case though the expression for the amplitude is
similar). Furthermore, we predict a change of sign at $Q^2 \approx
0.5-1$ GeV$^2$, a feature suggested by the present data. 
The behaviour of our amplitude resembles the one
obtained with a light-front formalism \cite{CARDARELLI97,CAPSTICK95},
though we do not have a significant configuration mixing as it was the
case in ref. \cite{CARDARELLI97}.

	In order to have a deeper understanding of the nature of the
Roper a 
test based on the calculation of the longitudinal amplitudes was proposed
\cite{LI92}. In figs. 3.c and 3.d we show our results for $S_{1/2}$. The
interpretation of experimental points \cite{GERHARDT80} in terms of
the longitudinal amplitude is controversial (see for instance refs.
\cite{CARDARELLI97} and \cite{LI92}) about the sign of this amplitude.
Then no definitive conclusion shoud be extracted before more confident
data are available and agreement about its interpretation is reached.
Meantime we may conclude that 
a $3q$ structure of the Roper cannot be discarded once the transition 
operator incorporates the physically relevant ingredients.

	The resulting picture of the Roper is that of a small quark
core dressed by a mesonic cloud of total charge zero. The
contribution of the cloud to the baryon masses can be taken into account
quite approximately, through the values of the effective parameters or
the potential, in 
a $3q$ state. For the transitions, the cloud contribution has to be
considered explicitely (as an explicit $|qqq \; q\bar{q} \rangle$
component or in the transition operator as we have done). 
No exotic explanation of the
Roper nature seems to be needed, reinforcing the similar conclusions
obtained in different frameworks \cite{CARDARELLI97,CAPSTICK95}.

	The analysis can be extended to other resonances
\cite{CANOPREP}. In particular, for the $\Delta(1600)$, the Roper of
the $\Delta$, the $Q^2$ behaviour of the amplitude is depicted in fig.
4, showing a much more pronounced quark model dependence. 
In our model it also corresponds to a radial excitation so the
comparison to data when available may be a more stringent test about
our conclusions. 

\vskip 1cm     

	We are grateful to S. Singh and S. Noguera for their
suggestions and comments and to  G. Salm\`e for his immediate answer
and clarification to our questions on the subject.
This work has been partially supported by DGES
under grants PB95-1096, PB94-0080 and by EC-TMR network
HaPHEEP under contract ERBTMRX CT96-0008.


\newpage 

\begin{center}
{\bf Table caption}
\end{center}

\begin{description}

\item{\bf Table 1} Charge radii (in fm$^2$) and magnetic moments (in
nuclear magnetons) for the nucleon in the two models $\vI$ and $\vII$.
The contribution from each diagram in fig. 2 is shown. The
photoproduction amplitudes $A_{1/2}^{p,n}$, $S_{1/2}^{p,n}$ (in units
of $10^{-3}$ GeV$^{-1/2}$) correspond to the process $N(1440)
\rightarrow N \gamma$. The EEM contribution includes the relative
weight $c$. 

\end{description}

\begin{center}
{\bf Figure captions}
\end{center}

\begin{description}

\item{\bf Figure 1} Extended vector meson dominance picture.

\item{\bf Figure 2.} Magnetic form factor for the proton calculated
in the EVMD model for the two potentials $\vI$ (solid line), $\vII$
(dashed line). For exp. data see for instance \cite{GMPDATA}

\item{\bf Figure 3.} Transverse (a,b) and longitudinal (c,d)
electroproduction amplitudes for N(1440). The solid line corresponds to
$\vI$ and the dashed line to  $\vII$. 
Experimental points at $Q^2=0$ in a) and b) 
are taken from \cite{PDG96}. For other experimental points see \cite{LI92}. 

\item{\bf Figure 4.} Transverse electroproduction amplitude for the
process \linebreak $\Delta^+(1600) \rightarrow p \gamma$. Notation as in fig. 3.

\end{description}

\newpage

\
\vfill
\begin{table}[h]
\begin{center}
{\tabcolsep 3pt 
\begin{tabular}{|c|rrr@{\hspace{2.ex}}|rrr@{\hspace{2.ex}}|c|}
\hline \hline
\rule{0pt}{3.ex}
& \multicolumn{3}{|c|}{$\vI$} & 
 \multicolumn{3}{|c|}{$\vII$} &  \\[2ex]
\cline{2-7}
\rule{0pt}{3.ex} &
{\small EEM} & {\small VMD} & {\small Total} & 
{\small EEM} & {\small VMD} & {\small Total} & 
\rule{0pt}{-15.ex} Exp. \\[2ex]
\hline  
\rule{0pt}{4ex}
$\langle r_p^2 \rangle $ & 
 0.238 &  
0.302 &
0.54 &
 0.133 &
0.407 &
0.54 & 
{\tabcolsep 1.pt 
\begin{tabular}{rcl}
0.74 & $\pm$ & 0.02
\end{tabular}} \\[1ex]
\rule{0pt}{2ex}
$\langle r_n^2 \rangle $ & 
$-0.02 $  &
$-0.012 $ &
$-0.032 $ &
$-0.005 $  &
$-0.01 $ &
$-0.015 $  &
{\tabcolsep 1.pt 
\begin{tabular}{rcl}
$-0.119$ & $ \pm$ & 0.004 
\end{tabular}} \\[1.25ex]
\hline
\rule{0pt}{3ex}
$ \mu_p $ &
0.93  &
1.86 &
2.79 &
0.27 &
2.52 &
2.79 & 
{\tabcolsep 1.pt 
\begin{tabular}{rcl}
 & 2.79 & 
\end{tabular}} \\[1ex]
\rule{0pt}{2ex}
$ \mu_n $ &
$-0.61$  &
$-1.21$ &
$-1.82$ &
$-0.18$ &
$-1.66$ &
$-1.84$ &
{\tabcolsep 1.pt 
\begin{tabular}{rcl}
 & $-1.91$ & 
\end{tabular}} \\[1.25ex]
\hline 
\rule{0pt}{3ex}
$ A_{1/2}^p $ &
0.63 &
$-82.1$ &
$-81.5$ & 
$-1.9 $ &
$-164 $ &
$-166 $ &
{\tabcolsep 1.pt 
\begin{tabular}{rcl}
$-65$ & $\pm$ & 4 
\end{tabular}}\\[1ex]
\rule{0pt}{2ex}
$ A_{1/2}^n $ &
$-.26 $  &
$52.3 $ &
$52.0$ & 
$1.3 $ &
$108 $ &
$110 $ &
{\tabcolsep 1.pt 
\begin{tabular}{rcl}
40 & $\pm$ & 10 
\end{tabular}} \\[1.25ex]
\hline 
\rule{0pt}{3ex}
$ - S_{1/2}^p $ &
$- 16.5 $&
$-30.3$ & 
$-46.8 $ &
$-4.7 $ &
$-61.4 $ &
$-66.1 $ &
{\tabcolsep 1.pt 
\begin{tabular}{rcl}
 & -- &  
\end{tabular}}\\[1ex]
\rule{0pt}{2ex}
$ - S_{1/2}^n $ &
2.1  &
3.7 &
5.9 & 
0.25 &
2.6 &
2.8 &
{\tabcolsep 1.pt 
\begin{tabular}{rcl}
& -- &  
\end{tabular}} \\[2ex]
\hline\hline	
\end{tabular}
}
\vskip 1cm
{\bf Table 1}
\end{center}
\end{table}
\vfill

\newpage

\begin{center}
\begin{tabular}{c}
\psfig{file=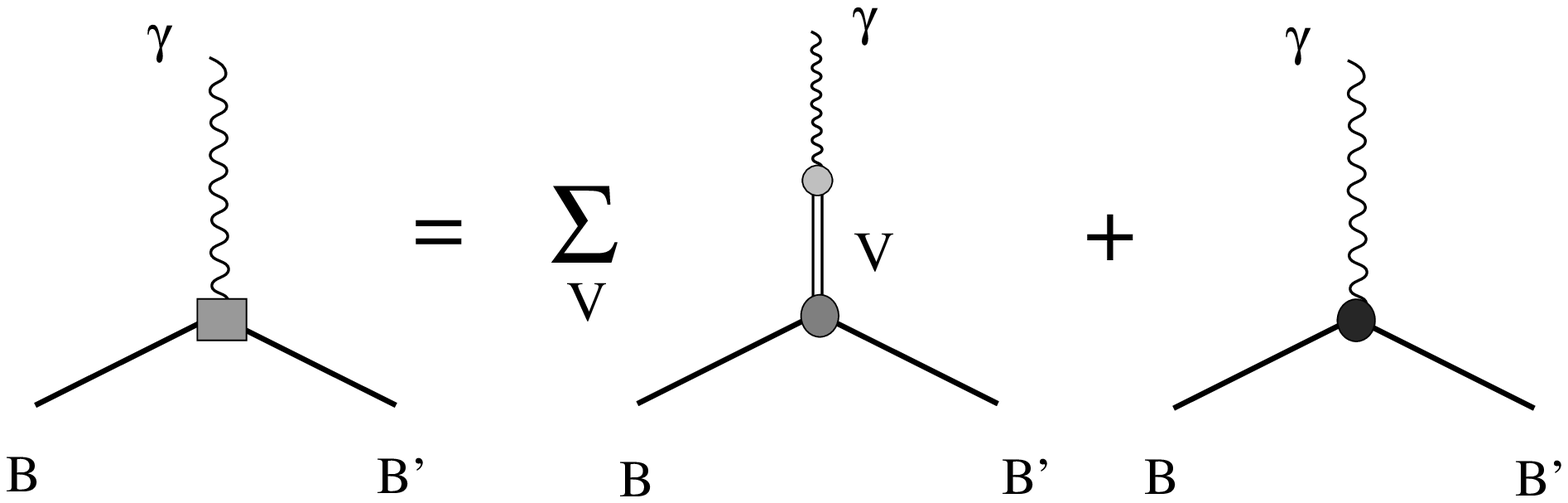,width=0.55\textwidth} \\
\rule{0pt}{8.ex}{\bf fig. 1} \\[35.ex]
\psfig{file=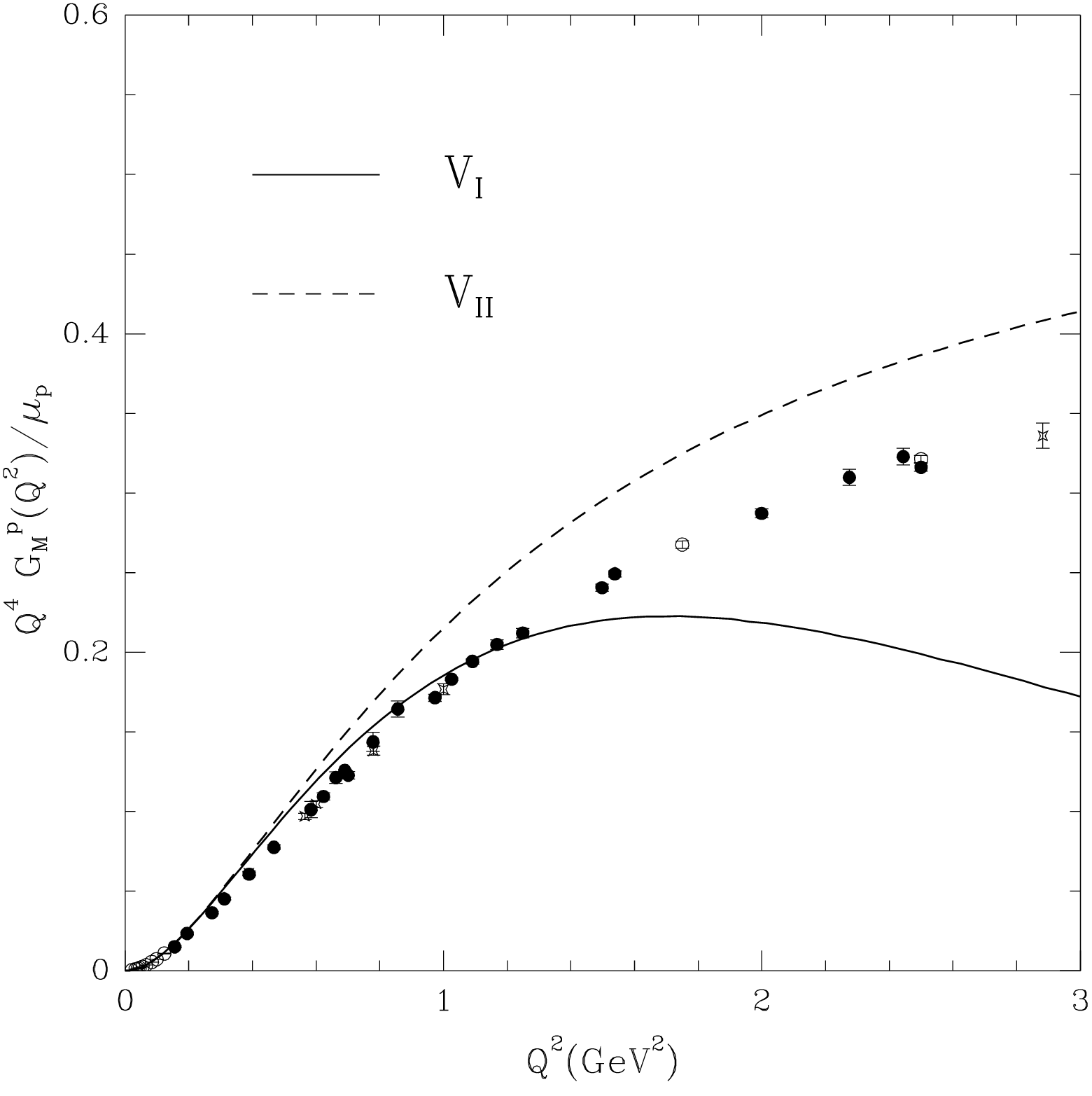,width=0.55\textwidth} \\
\rule{0pt}{8.ex}{\bf fig. 2}
\end{tabular}
\end{center}

\newpage

\begin{tabular}{cc}
\begin{tabular}{c}
\psfig{file=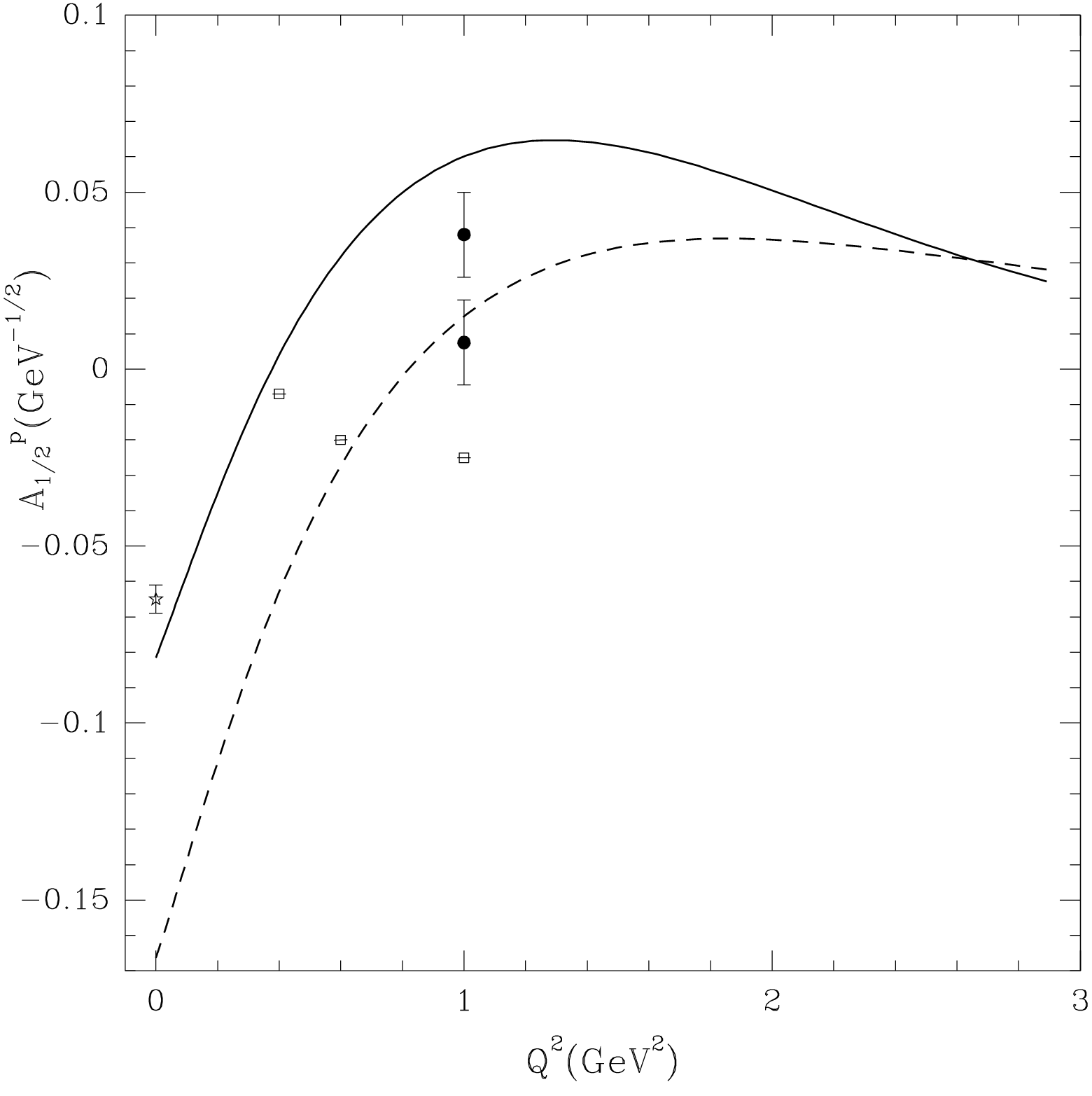,width=0.5\textwidth} \\
\rule{0pt}{8.ex}{\bf fig. 3a}
\end{tabular} & 
\begin{tabular}{c}
\psfig{file=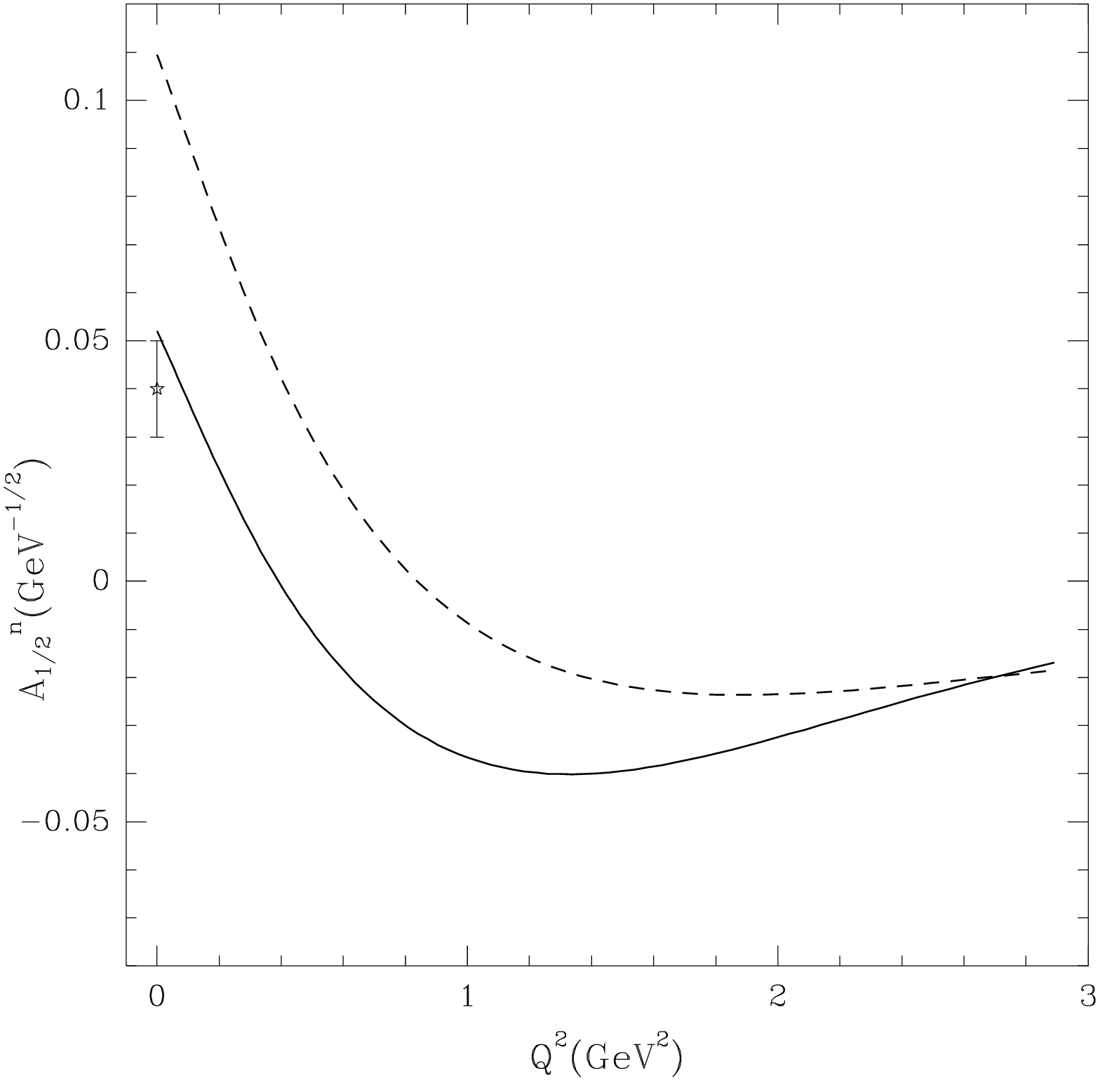,width=0.5\textwidth} \\
\rule{0pt}{8.ex}{\bf fig. 3b}
\end{tabular} \\[35.ex]
\begin{tabular}{c}
\psfig{file=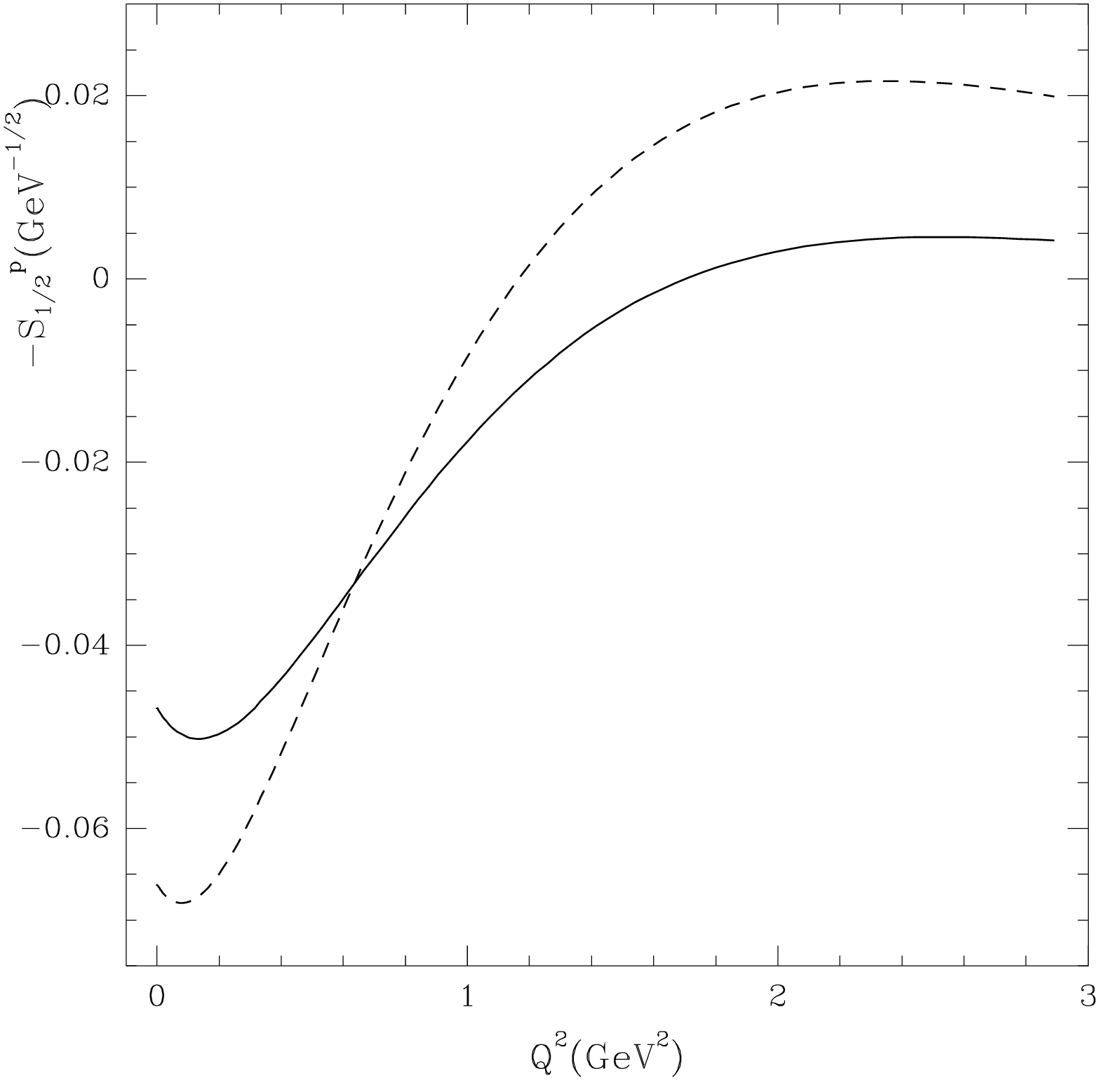,width=0.5\textwidth} \\
\rule{0pt}{8.ex}{\bf fig. 3c}
\end{tabular} &
\begin{tabular}{c}
\psfig{file=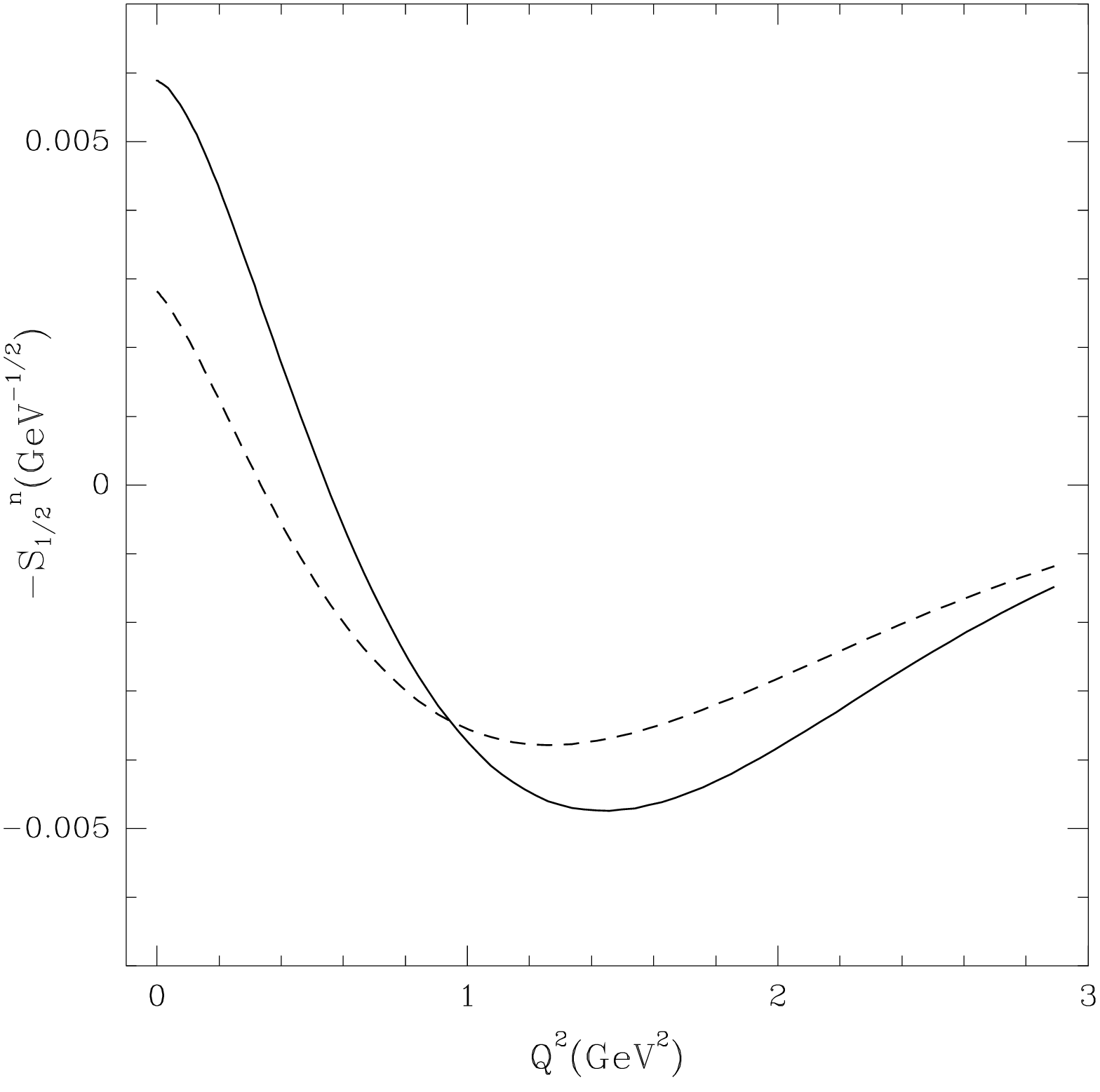,width=0.5\textwidth} \\
\rule{0pt}{8.ex}{\bf fig. 3d}
\end{tabular} 
\end{tabular}

\newpage

\
\vfill
\begin{center}
\begin{tabular}{c}
\psfig{file=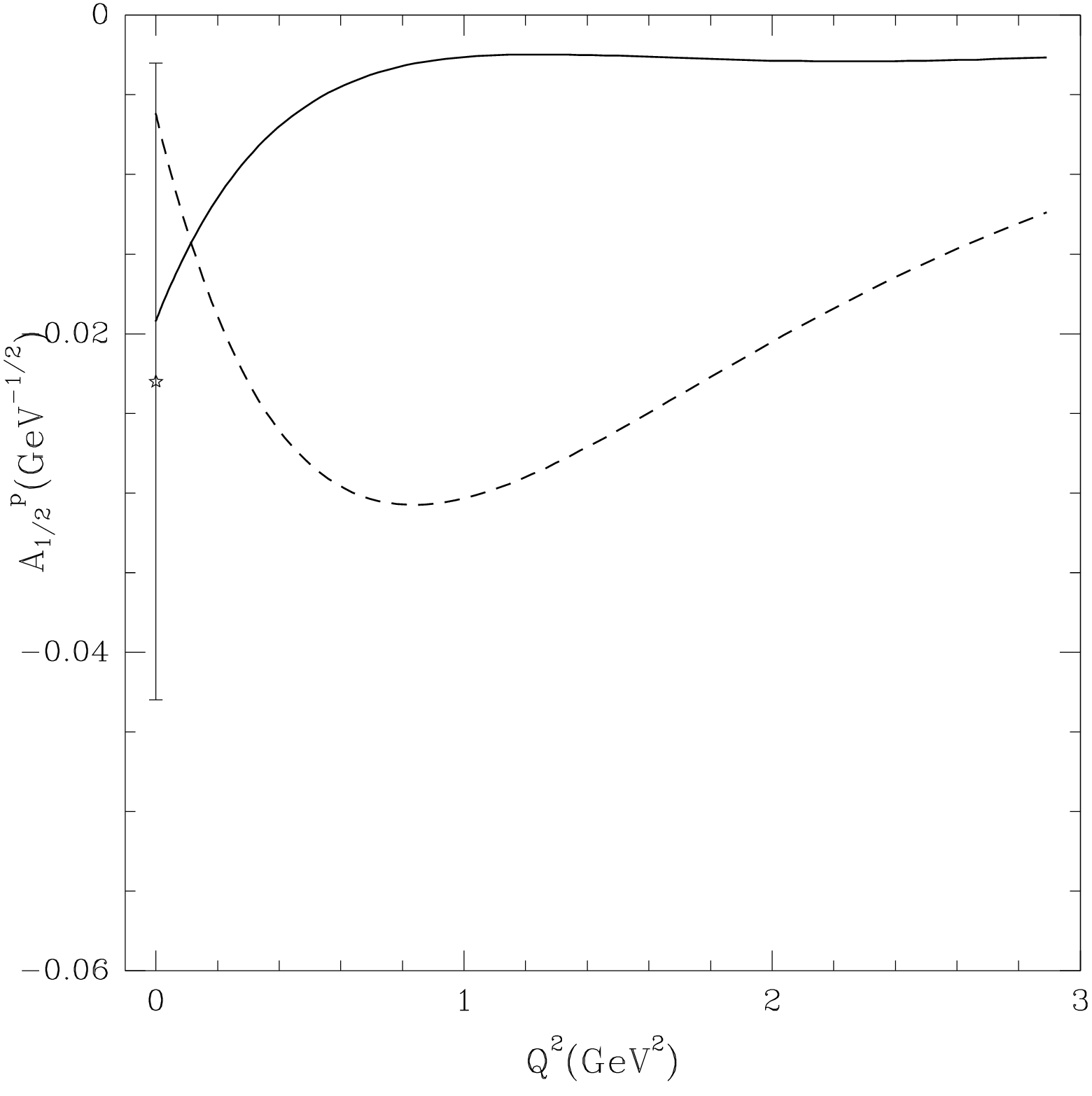,width=0.55\textwidth} \\
\rule{0pt}{8.ex}{\bf fig. 4}
\end{tabular} 
\end{center}
\vfill
\end{document}